\documentstyle[general_cite]{mn}

\title{An observational constraint on gravitational lensing by objects of mass
$10^{9.5}$--$10^{10.9}$~M$_{\odot}$} 
\author[Pedro Augusto and Peter N. Wilkinson]{Pedro~Augusto,$^{1,2}$ and 
Peter~N. Wilkinson,$^1$ \\
$^1$University of Manchester, Jodrell Bank Observatory, Macclesfield,
Cheshire SK11 9DL, UK.\\ $^2$Universidade da
Madeira, Dep.\ Matem\'atica, Caminho da Penteada, 9050 Funchal, Portugal\\}
\begin{document}
\bibliographystyle{mnras}
\maketitle
\begin{abstract}
A radio-based search for strong gravitational lensing, with image
separations in the range 160--300 milliarcsec (mas), has yielded a null
result for a sample of 1665 sources (Augusto, Wilkinson \& Browne
1998) whose mean redshift is estimated to be $\sim 1.3$. The lensing
rate for this previously-unexplored separation range, $<1:555$ at the
95\% confidence level, is less than on arcsecond-scales---as expected
from models of lensing galaxy populations. Lensing on 160--300 mas scales is
expected to arise predominantly from spiral galaxies at a rate
dependent on the disk-halo mass ratio and the evolving number density
of the population with redshift. While the present sample is too small
for there to be a high probability of finding spiral galaxy lenses,
our work is a pilot survey for a  much larger search based on the full CLASS database which would
provide useful information on galactic structure at $z \sim 0.5$. We
examine other possible lens populations relevant to our present search, in
particular dwarf galaxies and supermassive black holes in galactic
nuclei, and conclude that none of them are likely to be detected. Our
null result enables us formally to rule out a cosmologically
significant population of uniformly-distributed  compact objects:
$\Omega_{\rm CO}<0.1$ (95\% confidence) in the mass range
$10^{9.5}$--$10^{10.9}$~M$_{\odot}$.
\end{abstract}

 
\section{Introduction}

Systematic radio-based surveys have proved to be a very successful way
of finding gravitational lens systems with image separations in the
range 0.3--6 arcsec; the lensing masses are the central regions of
normal galaxies.  The Jodrell-VLA Astrometric Survey (JVAS)/Cosmic
Lens All-Sky Survey (CLASS) surveyed over 12000 sources, using the
VLA, MERLIN and the VLBA as successive high resolution filters. These
have, so far, found 19 such lens systems (Browne et al.\ 2000 and
unpublished). In the course of a general investigation of the
properties of flat-spectrum radio sources, drawn principally from the
JVAS, Augusto, Wilkinson \& Browne (1998; paper I hereafter) carried
out a search for multiple imaging with smaller separations in the
range $\sim$90--300 mas. Using essentially the same VLA, MERLIN and
VLBA filtering process employed in the main JVAS/CLASS lens surveys,
no cases of multiple imaging were found in a sample of 1665
sources. In this paper we examine the implications of this null
result.

The classic statistical study of gravitational lensing probabilities
was carried out by Turner, Ostriker \& Gott (1984; TOG84 hereafter)
who calculated the distribution of the angular separations of multiple
images expected from normal galaxies. TOG84 predicted that $\sim 88\%$
of lenses associated with normal galaxies should occur in the range
0.3--6 arcsec. All of the known cases of strong lensing by individual
galaxies fall in this angular size range, with a mean separation
consistent with TOG84's expectation of $\sim 1.5$ arcsec (Browne et
al.\ 2000) and a smallest separation of 0.335 arcsec \cite{Patetal93}.
 It is important, however, to subject the
TOG84 predictions to further observational scrutiny in order to see
whether there are other contributions to the lens population. Our
results are for a previously unexplored separation range and hence
provide a test for lower-mass galaxies (or components of galaxies)
than were considered by TOG84. This is particularly relevant since
recent calculations of the lensing contribution of spiral galaxies
suggest that TOG84 may have underestimated their contribution to the
lensing optical depth by more than a factor two (see section 2.1).

Our results also enable us to place observational constraints on
`non-standard' populations of potential lensing masses. Any mass
concentration with size $\la0.02 \sqrt{M/M_{\odot}}$ pc is `compact'
as regards gravitational lensing \cite{PreGun73} and unseen dark
Compact Objects (CO) could produce lensing on scales of interest for
this paper. If present in large enough numbers, such CO could
contribute to the solution of the dark matter problem.  In the extreme
case of a universe filled with a critical density of CO of a
particular mass ($\Omega_{\rm CO}=1$), the odds are even that any
object at $z=2$ will be multiply imaged into two images with a
brightness ratio $<10:1$ \cite{PreGun73}. The lensing probability
scales directly with the number density of CO and hence the statistics
of gravitational lensing in surveys of distant sources provide direct
constraints on $\Omega_{\rm CO}$.  Throughout this paper we take the
value of $H_0$ to be 65 km s$^{-1}$ Mpc$^{-1}$.

\section{Lenses of mass $\sim10^{9.5}$--$10^{10.9}$ M$_{\odot}$}

\begin{figure} 
\setlength{\unitlength}{1cm} 
\begin{picture}(8.0,6.0) 
\put(8.3,5.6){\includegraphics{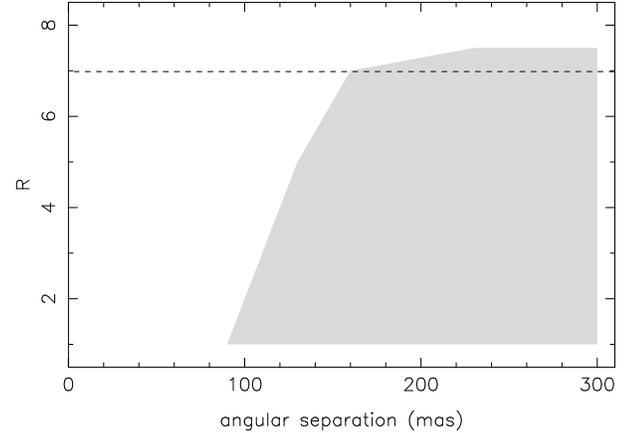}} 
\end{picture} 
\caption{The angular separations ($\delta$) of the multiple compact
components whose maximum flux density ratio is $R$, which will be
selected from our VLA A-array `snapshot' data at 8.4 GHz using the
criterion of a $\sim$ 25\% decrease in correlated flux density from
the shortest to the longest baselines. In the shaded region lens
candidates will be picked out with near 100\% reliability (reproduced
from paper I). In order to simplify our analysis we only considered
the range $160\leq \delta \leq 300$ mas where we are confident of
selecting sources with $R\leq7$.}
\label{papIfig} 
\end{figure}

Our basic observational result is that amongst a parent sample of 1665
flat-spectrum radio sources there are no cases of gravitational
lensing with separations in the range $\sim$90--300 mas. The only weak
candidate left from paper I, B0225+187, was ruled out from a
subsequent MERLIN+EVN 1.6 GHz map (Augusto, unpublished). For our
search, the flux-ratio ($R$) of candidate images is dependent on their
angular separation ($\delta$) since this is comparable to the beam
size of the VLA in A configuration at 8.4 GHz ($\sim 200$ mas).
Fig.~\ref{papIfig}, reproduced from paper I and discussed in detail
there, shows the angular selection function for our search derived
from extensive simulations with the VLA `snapshot' aperture-plane
coverage and a range of source types, position angles and
declinations. We were conservative in deriving this selection
function, which represents the `lensing filter' at essentially 100\%
confidence. Sources with smaller separations could be detected, but
less often, depending on their position angle. For example, an equal
double source ($R=1$) with a separation of 50 mas had a $\sim 50\%$
chance of being included and one with 75 mas separation had
$\sim97\%$.  However, in order to simplify our analysis and its
interpretation, we will only consider separations in the range
160--300 mas where we are confident (Fig.~1) of selecting sources with
$R \sim 7$---the usual definition of strong lensing (e.g. TOG84).

We can, therefore, place a formal upper limit on the lensing rate in
the range 160--300 mas using the equation
\begin{equation} 
P(0,N)=(1-p)^{N},
\label{prob} 
\end{equation} 
where $P(0,N)$ is the probability of finding no cases of multiple
imaging amongst $N$ sources (1665 in our case) when the probability of
lensing an individual source is $p$. To obtain an upper limit to $p$
at the 95\% confidence level we require $P(0,N) \leq 0.05$ and hence
obtain $p=0.00180$. The upper limit ($1/p$) on the strong-lensing rate
for image separations 160--300 mas is, therefore, $1:555$ (95\%
confidence). We now examine the implications of this result for
various possible lensing objects.

\subsection{Normal galaxies}

Since the parent sample for our lens search is dominated by JVAS
sources \cite{Aug96} we can directly compare our null result with the
lensing rate at arcsec-scale separations found in 2384 JVAS sources
\cite{Kinetal99}. Since five suitable\footnote{One of the JVAS lenses
(B1030+074) has $R \sim 15$ and hence has not been included in our
estimate of the lensing rate which is based on JVAS lenses with $R
\leq 7$.} JVAS lenses were found, the rate is $\sim 1:478$. Our null
result shows that, as predicted by TOG84, the lensing rate is almost
certainly lower for image separations of $\sim 0.2$ arcsec than it is
for arcsec-scale separations. Taking the argument one step further,
from TOG84's probability distribution we estimate that only $\sim 6\%$
of lensing by normal galaxies should produce image separations in the
range 160--300 mas, i.e., the lensing rate should be about 1:7000.

The dominant contribution to the overall gravitational lensing
cross-section in TOG84's models is by elliptical galaxies; spiral
galaxies are expected to contribute only $\sim 20\%$ of the
total. This is consistent with the known population of arcsec-scale
lenses which only contain a few spirals. However, since TOG84
published their paper, there has been considerable further work on the
lensing properties of disks embedded in dark matter halos
(e.g. \pcite{KeeKoc98,BarLoe98}; Blain, M\"{o}ller \& Maller 1999;
\pcite{Bar00}).  The cross-sections are dominated by edge-on disks and
the predicted lensing rate is dependent on the balance between the
disk and halo masses. \scite{KeeKoc98} predict that, when averaged
over all inclinations, there should be little change in the
contribution of spirals compared with the predictions of TOG84. In
contrast the other models \cite{BlaMolMal99,BarLoe98,Bar00}, which
involve maximal disks and also consider the effect of evolution of the
spiral population with redshift, predict significant enhancements of
the total spiral fraction, by factors of two or more, compared with
TOG84.

Since spirals contribute most at small image separations, the
enhancement over TOG84 could become large for the previously
unexplored separations to which our search is sensitive.
For example,
Bartelmann (priv.\ comm.) estimates that between 10\% and
20\% of {\it all} galaxy-mass lenses could have separations in the
range 0.1--0.3
  arcsec as
opposed to $\sim8\%$ in TOG84. These models suggest, therefore, that
the lensing rate appropriate to our survey is one in a few thousand
background objects searched. Thus our 1665-source parent sample still
does not have a high probability of containing a small-separation
spiral lens.  However, an order-of-magnitude larger search would place
significant contraints on the uncertain disk/halo mass ratio in spiral
galaxies at $z\sim 0.5$. We return to this point later.

\subsection{Dwarf  Galaxies}

TOG84 did not consider lensing by dwarf galaxies. However, our search is
sensitive to the upper end of the dwarf mass range and so we must ask
the question: are the properties of the dwarf galaxy population such
as to make multiple imaging likely in the present context? There are
two issues to address: (i) are dwarf galaxies compact enough to have
the critical surface mass density ($\sim 1$~g cm$^{-2}$) for multiple
imaging? and (ii) are there enough of them to compensate for their
much smaller lensing cross-section compared with normal ellipticals?
We will address these issues briefly in turn.

The mass distributions of dwarf galaxies are in general poorly
known. However, from a study of published HI rotation curves of nine,
gas-rich, dwarf irregulars (dI) \scite{Aug96} showed that their mean
surface mass densities are one to two orders of magnitude less than
those of the normal ellipticals which dominate the lensing
statistics. On this basis dI can be ruled out as potential lenses. The
surface mass densities of gas-poor dwarf ellipticals (dE) are
essentially unknown but since most show compact nuclei (e.g.\
\pcite{BinCam91,FerSan89,VadSan91}) it is possible that dE do have the
requisite surface mass density for multiple imaging. The remaining
types of dwarf galaxies are a lot less common.

Better determinations of the space density of dwarf galaxies have
recently become available. For example, from studies of the Hubble
Deep Field, \scite{Dri99} shows that while dwarf galaxies at $0.3 < z
< 0.5$ (which are well-placed for lensing) are more numerous than
giant galaxies, the excess is only by factors of order
unity. Essentially the same conclusion is reached by \scite{Lov97}
from independent ground-based data. \scite{Dri99} goes on to show
that, for reasonable assumptions about their mass-to-light ratios,
dwarf galaxies only account for $\sim 16\%$ of the total galaxy mass
budget.

Since the separation range to which our search is sensitive is 5--10
times smaller than the typical separation of lens systems in JVAS the
corresponding lensing cross-section is 25--100 times smaller. We
conclude, therefore, that even if {\em all} dE are capable of multiple
imaging, on current estimates their numbers are too small, by at least
an order-of-magnitude, to produce a significant number of lensing
events in a sample of 1665 objects.

\subsection{Dark Compact Objects}

\subsubsection{Supermassive black holes in galaxies}

We can be reasonably certain of one population of dark compact objects
relevant to the present search---the supermassive black holes (SBH)
which power active galactic nuclei (AGN) and which may well lie hidden
in many inactive galactic nuclei (IGN). Quasars are most likely to
host a $\sim 10^9$ M$_{\odot}$ SBH but their average separation on the
sky for $z<1$ (appropriate for lensing) is $\sim 1000$ arcsec
(e.g. the 2dF survey at {\it www.mso.anu.edu.au/~rsmith/QSO\_Survey})
and hence, for an Einstein radius of $\sim 0.1$ arcsec, the lensing
probability for background sources is very low ($\sim10^{-8}$). Even
in the limit that all normal galaxies harbour `dead' quasars and
hence contain a SBH (e.g. Magorrian et al.\ 1998), and even if all these
SBHs have masses $\sim 10^9$ M$_{\odot}$, their associated lensing rate
must be two orders of magnitude below the JVAS rate for arcsec-scale
lensing simply because of the smaller cross-section.  We cannot,
therefore, place a useful constraint on the numbers of SBH in IGN from
our present results, nor from any future search of a ten-times larger
parent sample.

\subsubsection{Relic supermassive black holes}

It has long been been suggested (e.g. \scite{PreGun73}; Carr, Bond \&
Arnett (1984) and references therein) that a population of black holes
could have formed early in the history of the universe and could
remain in existence, perhaps providing seeds for AGN (e.g. \pcite{FukTur96}).  \scite{TurUme97} suggested
that relic SBH might be detected by means of occasional very strong
gravitational lensing of luminous stars in distant galaxies but, as
was first suggested by \scite{PreGun73} it is high resolution radio
imaging which allows the most direct search for lensing by relic SBH.
\scite{FukTur96} estimate that the number density of SBH of mass $\sim
10^9$~M$_{\odot}$ could be comparable with that of luminous galaxies
and up to three orders of magnitude higher than the peak number
density of quasars. But even so, from the simple cross-section
arguments, it is clear that much larger samples than the present one
will be needed to put the hypothesis of a relic SBH population to a
severe observational test.

\subsubsection{Dark galaxies}

\scite{Pee68} proposed that a large number of `dead galaxies' could
solve the dark matter problem while more recently it has been proposed
that very massive ($10^{12}$--$10^{13}$ M$_{\odot}$) dark objects could give
rise to the quasar pairs with separations $\sim 10$ arcsec by
gravitational lensing \cite{Haw97}. There is, however, no supporting
evidence for such a population of massive dark objects. 
Kochanek, Falco and Munoz (1999)  argue that a
comparison of the radio and optical properties of the pairs rules out
the massive lens hypothesis. Furthermore, HST imaging of confirmed
arcsec-scale lenses found in the JVAS/CLASS surveys always shows a
lensing galaxy with a relatively normal mass-to-light ratio between
the images \cite{Jacetal98}.

There is a strong increase with redshift of the fraction of
morphologically peculiar galaxies (see the review by \scite{Ell00}
and references therein). It seems most likely that many of the distant
blue irregular galaxies are transformed by mergers into normal
ellipticals and spirals. It is, however, possible that they could be
seen at an unusually active period in their history and that, starved
of infalling gas, they could have simply faded away to low surface
brightness (LSB) systems which would be hard to detect (e.g. \pcite{Ell00}). 
However, without a model of the remaining LSB systems fraction and
their likely mass distribution, one cannot make a prediction as to the
likelihood of detecting them by their lensing effects.

Although there is no evidence for them in arcsecond-scale lens
searches, dark galaxies with smaller masses of
$\sim10^{9.5}$--$10^{10.9}$ M$_{\odot}$ (and respective upper limits
on their sizes of $\sim1.2$--5.6 kpc) could exist and would produce
multiple images of background sources in the range 160--300 mas.  For
these we can place the same cosmological density limits as for compact
objects (e.g.\ black holes) and this is the subject of the next
section.

\subsection{Quantitative limits on $\Omega_{CO}$}

Despite the apparent lack of lensing objects which would produce
160--300 mas image separations with a significant probability in our
1665-source sample, it is still useful to set quantitative limits on
$\Omega_{CO}$ based on our null result. To do this we adopt the
`detection volume' method introduced by \scite{Nem89}.  Kassiola,
Kovner \& Blandford (1991) made specific calculations for
radio-surveys and hence these directly apply to our case. The
`detection volume' method consists of three steps, once a survey is
complete: i) derive the volume between each source in the survey and
the observer (`detection volume'); ii) add up the `detection volumes'
for all the sources in the survey; iii) compare the resultant total
volume with the volume expected for a given value of $\Omega_{CO}$.
 
In carrying out this calculation we make the following assumptions: i)
an Einstein-de-Sitter universe ($\Omega_{0}=1$; $\lambda_{0}=0$); ii)
no magnification bias; iii) $\Omega_{CO}\ll 1$. For our present
purpose assumption (i) is conservative. If the geometry of the
universe is in part determined by the cosmological constant, then the
expected number of lenses in a given sample rises (e.g.\
\pcite{Fuketal92}). A null lensing result therefore implies a stronger
limit on the number density of lensing objects on such a
Universe. Assumption (ii) is also conservative. Due to magnification
by gravitational lensing, flux-limited samples contain intrinsically
less luminous sources than the flux density threshold would normally
allow (see review on \pcite{NarWal93}). For flat-spectrum radio
sources, however, the magnification bias is expected to be relatively
weak \cite{KinBro96} and we have therefore not taken it into
account. If there is any bias our null result constitutes a stronger
constraint on $\Omega_{CO}$, since the actual number of sources in the
parent sample would increase. Assumption (iii) allows us to ignore
collective lensing effects of the sort explored by
\scite{KasKovBla91}.

Following \scite{KasKovBla91}, for a point mass lens and a background
source of a given redshift, we calculate the probability of {\em not}
detecting primary and secondary images with a flux ratio smaller than
$R$, the threshold at which the selection process stops reliably
picking lens candidates. The `detection volume' formulae presented in
\scite{KasKovBla91} are truncated because of the observational limits
of a survey. Thus there is a limit for the `resolution' of the
instrument used ($\delta$) and also for the maximum image separation
allowed for the lens candidates ($\Delta$). Following
\scite{KasKovBla91}, we have for the truncated detection volume:
\[ V_{t} = \frac{16\pi G M}{H_{0}^{2}} (R^{1/4} - R^{-1/4})^{2} \nu(z_{s},M,\delta,\Delta)  \] 
with $M$ the mass of the point lens, $z_s$ the redshift of the
background source and $\nu$  a dimensionless volume function. While
$\nu$ is a function of mass, $\delta$ and $\Delta$ effectively define
lower and upper limits of a mass range for which $\nu$ is
approximately constant and the truncation is small. Outside this mass
range truncation increases sharply and $\nu$ (and  $V_{t}$)
rapidly tends to zero. For the mass range $M_{1}<M<M_{2}$ the limit on
$\Omega_{CO}^{M_{1}-M_{2}}$ at the $100\Pi\%$ confidence level is
obtained from:
\begin{equation}  
\Omega_{CO}^{M_{1}-M_{2}} < \frac{- \ln (1-\Pi)}{6 \; (R^{1/4} -
R^{-1/4})^{2}  \sum_{i} \nu(z_{s_{i}},M,\delta,\Delta) }.
\label{limitOm} 
\end{equation} 

\begin{figure} 
\setlength{\unitlength}{1cm}
\begin{picture}(16.0,5.5) 
\put(15.4,-3.7){\includegraphics{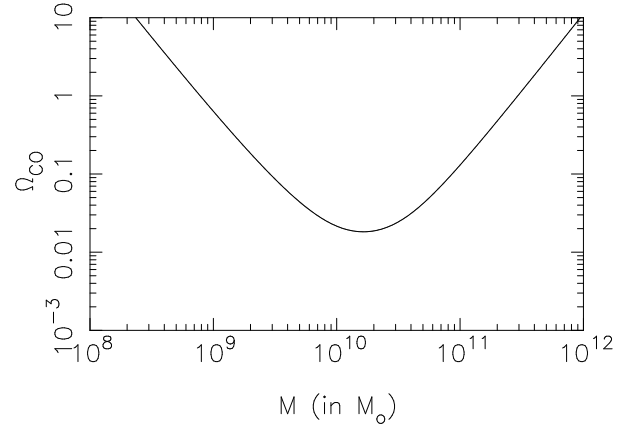}} 
\end{picture} 
\caption{The limits on $\Omega_{CO}$ at the 95\% confidence level from
the failure to detect multiple imaging amongst 1665 flat-spectrum sources with
a mean redshift of 1.3. These are $\Omega_{CO}^{9.5-10.9}<0.10$ (on the $\sim
10^{9.5}$--$10^{10.9}$~M$_{\odot}$ mass range), $\Omega_{CO}^{9.9-10.6}<0.03$
and $\Omega_{CO}^{10.2}<0.02$ (strongest constraint) for objects with mass 
$\sim10^{10.2}$~M$_{\odot}$.} 
\label{logP0_graph} 
\end{figure}

Our search parameters appropriate for equation~\ref{limitOm} are:
$\delta=160$~mas; $\Delta=300$ mas and $R=7$. These are conservative
parameters derived from the complicated $R(\delta)$ function of
Fig.~1. \scite{Aug96} has shown that the results obtained by including
the range 90--160 mas are similar.  To determine $\sum_i \nu$, we need
redshift information, but this is very limited for the $z_{s_{i}}$ of
the 1665-source parent sample. We can, however, make use of the 90\%
complete redshift information of the flux-limited 293-source CJF
sample \cite{Tayetal96}, which is drawn largely from the JVAS
catalogue.  The lower-limiting flux density of the CJF sources is 350
mJy at 5 GHz compared with $\sim 130$ mJy for our sample.  There is,
however, no significant variation in the redshift distributions and
the mean redshift ($\sim 1.3$) of flat-spectrum radio sources in the
range 1 Jy to $\sim30$ mJy (e.g. Falco, Kochanek \& Munoz 1998;
\pcite{Maretal00}) and so little error will arise from applying the
well-defined CJF statistics to our sample.  The formal constraints on
$\Omega_{CO}$ at the 95\% confidence level are then shown in
Figure~\ref{logP0_graph} and they are $\Omega_{CO}^{9.5-10.9}<0.10$.


\section{Discussion}

Among a sample of 1665 flat-spectrum radio sources, with an assumed
mean redshift of $\sim 1.3$, no multiple images were found with
separations in the range 160--300 mas (paper I and Augusto,
unpublished). The lensing rate for compact masses in the range $\sim
10^{9.5}$--$10^{10.9}$~M$_{\odot}$ is therefore less than $1:555$ at
95\% confidence. This result does not contradict TOG84 predictions
based on the properties of normal E/S galaxies.  No unexpected
population of small galaxies has thus been detected and it seems that,
despite their large numbers, $\sim10^9$ M$_{\odot}$ dwarf galaxies are
not contributing significantly to multiple imaging on 160-300
mas. This is consistent with the current knowledge about their numbers
and mass distributions.  

Recent work on the contribution of $z \sim 0.5$ spirals to the overall
lensing statistics gives contradictory results: there could be
significant (greater than a factor of two) enhancements of the total
spiral fractional contribution.  Part of the motivation for this new
work on the spiral population has been to anticipate the capabilities
of ALMA and NGST at $\sim$0.1 arcsec resolution in the sub-mm and
infra-red bands.  Radio-based surveys are similarly well-suited to an
unbiased search for spiral galaxy lensing since they are also not
affected by dust obscuration in edge-on disk systems. It is now
possible to mount a significant test of this prediction, since by
using the same methodology as adopted for Paper~I we can look for
compact lenses in the now-completed CLASS surveys.  Since the combined
JVAS/CLASS surveys have, so far, found 19 lenses with separations
0.3--6 arcsec in a total sample of $\sim12000$ sources, a search in
the range 0.1-0.3 arcsec using the same parent sample should find some
small-separation lenses (likely to be spirals). The results would
place significant contraints on the uncertain properties of spiral
galaxies at the redshifts appropriate to multiple imaging ($z\sim
0.5$).

Analysing the simpler point mass lens constraints, we conclude that
the cosmological density of compact objects  (which
include black holes and $\sim$1--5 kpc compact galaxies)
 in the mass range
$\sim3\times10^{9}$ to $\sim8\times10^{10}$ M$_{\odot}$, has to be
$\Omega_{CO}^{9.5-10.9}<0.10$ (95\% confidence level).
From our new result, it seems that $10^{9.5}$--$10^{10.9}$ M$_{\odot}$ SBH in this range cannot
contribute significantly to solving the dark matter problem.


\section*{Acknowledgments}
Pedro Augusto thanks the Funda\c c\~ao para a Ci\^encia e a Tecnologia
for the grant Ci\^{e}ncia/PraxisXXI/BD2623/93-RM. We acknowledge many
discussions with Ian Browne and Shude Mao and thank Matthias
Bartelmann for a specific discussion on spiral lens statistics.

\nocite{Magetal98,TurOstGot84,papI,BlaMolMal99,CarBonArn84,FalKocMun98,KocFalMun99,Broetal00}

\bibliography{limits_paper}

\end{document}